\documentclass{article}

\usepackage{arxiv}
\usepackage[utf8]{inputenc} 
\usepackage[T1]{fontenc}    
\usepackage{hyperref}       
\usepackage{url}            
\usepackage{booktabs}       
\usepackage{amsfonts}       
\usepackage{nicefrac}       
\usepackage{microtype}      
\usepackage{mathrsfs}
\usepackage{graphicx}
\graphicspath{ {./images/} }

\usepackage{tabularx}
\usepackage{ltablex} 
\usepackage{standalone}
\usepackage{graphicx} 
\usepackage{comment}
\usepackage{epigraph}
\usepackage{amsmath}
\usepackage{amsfonts}
\usepackage{amssymb}
\usepackage{cite}
\usepackage{caption}
\usepackage{subcaption}
\usepackage{subcaption}
\usepackage{latexcolors}
\usepackage{import}
\usepackage{tikz}
\usetikzlibrary{positioning}
\usetikzlibrary {arrows.meta}
\usepackage[ruled,vlined]{algorithm2e}

\usepackage{mathtools}
\mathtoolsset{showonlyrefs}

\newcommand{\genecard}[1]{\href{https://www.genecards.org/cgi-bin/carddisp.pl?gene=#1}{#1}}

\title{friends.test: rank-based method for feature selection in interaction matrices}

\author{
  Alexandra Suvorikova \thanks{These authors contributed equally to this work.} \\
  Weierstrass Institute;\\
  Institute for Information\\Transmission Problems RAS\\
  \texttt{suvorikova@wias-berlin.de} \\
  \And
  Alexey Kroshnin\footnotemark[1] \\
  Weierstrass Institute\\
  \And
  Dmirijs Lvovs \\
  Institute for Genome\\Sciences,\\University of Maryland\\School of Medicine\\
  \And
  Vera Mukhina \\
  Vavilov Institute for\\General Genetics RAS\\
  \AND
  Andrey Mironov \\
  Faculty of\\Bioengineering and\\
  Bioinformatics MSU
  \And
  Elana J. Fertig \\
  Institute for Genome\\Sciences and Greenbaum\\Comprehensive Cancer Center,\\University of Maryland\\School of Medicine
  \And
  Ludmila Danilova \\
  Johns Hopkins University\\School of Medicine \\
  \And
  Alexander Favorov \\
  Johns Hopkins University\\School of Medicine;\\Vavilov Institute for\\General Genetics RAS \\
  \texttt{favorov@sensi.org} \\
}

\begin{document}
\maketitle
\begin{abstract}
The analysis of the interaction matrix between two distinct sets is essential across diverse fields, from pharmacovigilance to transcriptomics. Not all interactions are equally informative: a marker gene associated with a few specific biological processes is more informative than a highly expressed non-specific gene associated with most observed processes. Identifying these interactions is challenging due to background connections. Furthermore, data heterogeneity across sources precludes universal identification criteria.

To address this challenge, we introduce \textsf{friends.test}, a method for identifying specificity by detecting structural breaks in entity interactions. Rank-based representation of the interaction matrix ensures invariance to heterogeneous data and allows for integrating data from diverse sources. To automatically locate the boundary between specific interactions and background activity, we employ model fitting. We demonstrate the applicability of \textsf{friends.test} on the GSE112026---transnational data from head and neck cancer. A computationally efficient \textsf{R} implementation is available at \url{https://github.com/favorov/friends.test}.
\end{abstract}


\section{Introduction}

Many modern problems involve understanding the interaction between two sets of objects. For instance, recommendation systems link users to movies, pharmacovigilance connects drugs to adverse effects \cite{timilsina2019discovering}, and transcriptomics associates genes with biological processes \cite{fertig_cogaps_2010, stein-obrien_enter_2018}. However, not all interactions are equally informative. For instance, in the analysis of protein-protein interaction maps, prioritizing proteins that interact strongly with only a narrow set of biological processes---rather than those with broad, non-specific connectivity---can improve therapeutic specificity and reduce off-target effects \cite{viacava2021centrality}. Genes that are uniformly expressed across all samples do not contribute to the identification of specific biological states. Instead, the analysis relies on tissue-specific markers that provide a clear signal for differentiating unique processes
\cite{fertig_cogaps_2010}. 

Selecting entities for analysis solely by their interaction strength often fails to distinguish true relevance from non-specific background activity, because in many domains meaningful functional relationships are confined to a narrow subset of interactions. For instance, modern genetic studies highlight the necessity of quantifying gene specificity to better identify signals unique to a particular trait \cite{spence2025specificity}. This phenomenon reflects a fundamental challenge in the analysis of bipartite interaction data. 

In this work, we assume the data are represented by an interaction matrix $A$ of size $n \times k$, where rows correspond to a set of entities $T = \{t_1, \dots, t_n\}$ (e.g., genes) and columns denote a set of counterparts $C = \{c_1, \dots, c_k\}$ (e.g., biological processes):
\begin{equation}
\label{eq:adj_matrix}
A := \begin{pmatrix}
a_{11} & a_{12} & \dots & a_{1k} \\
 &\cdots & \cdots & \\
a_{n1} & a_{n2} & \dots & a_{nk}
\end{pmatrix}.
\end{equation}
Each entry $a_{ij}$ represents the strength of interaction between $t_i$ and $c_j$. 
We refer to the $i$-th row vector $\mathrm{row}_{i}(A) = (a_{i1}, \dots, a_{ik})$ as the interaction profile of $t_i$, which encapsulates its connectivity pattern across all counterparts in $C$. 

We model the presence of informative interactions through a specific configuration of the interaction profile, which we term ``friendship''. We assume that an entity (row) $t_i \in T$ does not necessarily interact with its counterparts in $C$ uniformly; instead, it may exhibit selective affinity toward a specific subset of ``friends''. That is, its profile exhibits a clear transition (a structural break) between a subset of high-intensity interactions and a broader set of non-specific, background activity. For example, a gene might be selectively expressed in only a few specific biological processes while remaining at baseline levels elsewhere.

However, identifying such ``friends'' remains challenging for two main reasons. First, the experimental data often originate from heterogeneous sources or lack a common scale, making direct comparison of interaction strength non-informative---for example, when comparing gene expression levels across different processes or disparate experimental conditions. Second, it is unknown \textit{a priori} whether an entity exhibits ``friendship'' behavior at all. Even when such a pattern exists, neither the size of the high-intensity subset nor the magnitude of the structural break is known. Consequently, any threshold used to separate ``friendship'' from background activity must be inherently adaptive. 

To address these challenges, we introduce \textsf{friends.test}: a computationally efficient, self-tuning approach for detecting specific interactions by identifying structural breaks in interaction profiles. To overcome the lack of a common scale, our method utilizes a rank-based representation, which normalizes disparate interaction strengths and ensures the procedure is scale-invariant. Additionally, we employ model fitting to make the method adaptive; this allows the algorithm to automatically locate structural breaks and determine entity-specific thresholds, effectively distinguishing meaningful signals from background noise for each entity.

To demonstrate the utility of the \textsf{friends.test}, we apply it to a transcriptomic dataset of head and neck squamous cell carcinoma (HNSCC). \textsf{R}-package is available at \url{https://github.com/favorov/friends.test}. The package 
 runs in $\mathcal{O}(nk\log(n))$ times, where $n$ is the number of rows, $k$ is the number of columns. That performance makes the package scalable for large matrices. 

The paper is organized as follows. Section~\ref{sec:method} introduces the methodology. Section~\ref{sec:experiments} provides experimental results. Section~\ref{sec:discussion} discusses the algorithm---its limitations, possible applications, and related works.

\paragraph{Accepted notations.} We consider a dataset represented by an interaction matrix $A$ of size $n \times k$, where the rows correspond to a set of entities $T = \{t_1, \dots, t_n\}$ (e.g., genes or users) and the columns correspond to a set of objects $C = \{c_1, \dots, c_k\}$ (e.g., biological processes or movies).
For any index pair $(i, j)$, let $a_{ij}$ denote the observed interaction strength between entity $t_i$ and object $c_j$. We denote the $i$-th row of the matrix as $\mathrm{row}_i(A) := (a_{i1}, \dots, a_{ik})$ and the $j$-th column as $\mathrm{col}_j(A) := (a_{1j}, \dots, a_{nj})$. Throughout the paper, we use $F_i \subset C$ to denote the specific subset of ``friends'' for entity $t_i$. Finally, $\mathcal{U}\{\cdots\}$ denotes the uniform distribution.

\section{Methodology}
\label{sec:method}
The procedure is divided into three logical stages: normalizing the data to ensure scale-invariance, formalizing the structural break, and applying a decision rule to distinguish ``friendship'' from background noise. Algorithm~\ref{alg:friends} summarizes the procedure.

\subsection{Scale-invariant data representation}
We assume that each column in $A$ may follow its own scale or distribution. To model this effect, we introduce a latent variable framework. Let $\xi_{ij}$ ($1 \le i \le n, 1 \le j \le k$) be latent random variables. For each column $c_j \in C$, we assume there exists a fixed unknown strictly monotone increasing function $f_j: \mathbb{R} \to \mathbb{R}$ such that each observed interaction strength is given by $a_{ij} = f_j(\xi_{ij})$. Under this assumption, higher values of $a_{ij}$ indicate a stronger underlying interaction between $t_i$ and $c_j$.To distinguish between ``friendly'' and background interactions, we model the latent variables $\xi_{ij}$ as being drawn from an unkown two-component mixture, $\xi_{ij} \sim \pi \cdot P_{\text{friend}} + (1 - \pi) \cdot P_{\text{noise}},$
where $P_{\text{friend}}$ and $P_{\text{noise}}$ represent the distributions of ``friendly'' and non-informative interactions, respectively, and $\pi \in [0, 1)$ is the mixture weight.

Since $f_j$ is strictly monotone, it preserves the relative ordering of elements within each column $\text{col}_j(A)$. Consequently, converting the observed values $a_{ij}$ to ranks eliminates the unknown column-specific distortions $f_j$. This rank-based transformation recovers the underlying ordinal structure of the latent signals $\xi_{ij}$, thereby making the columns statistically comparable.

So, for each column $c_j \in C$, we rank the entries in $\mathrm{col}_j(A)$ in decreasing order, assigning the top rank to the largest value $a_{ij}$. In cases where multiple entries in $\mathrm{col}_j(A)$ share the same value, we use a randomized tie-breaking procedure. 

We denote the matrix containing the obtained ranks $r_{ij}$ as $R$, 
\begin{equation*}
R = \begin{pmatrix}
r_{11} & r_{12} & \dots & r_{1k} \\
 &\cdots & \cdots & \\
r_{n1} & r_{n2} & \dots & r_{nk}
\end{pmatrix}, 
\quad
r_{ij} :=\mathrm{rank}\left(a_{ij}~ \text{inside}~\mathrm{col}_j(A)\right).
\end{equation*}
We refer to $\mathrm{row}_i(R) = (r_{i1}, \dots, r_{ik})$ as normalized interaction profile of $t_i\in T$.

\subsection{The structural break model}
To identify ``friends'' of $t_i$, we model the normalized interaction profile $\mathrm{row}_i(R) = (r_{i1}, \dots, r_{ik})$ using a mixture of two uniform distributions on a discrete grid. For simplicity, we omit the index $i$ and denote the corresponding ranks as $r_{1}, \dots, r_{k}$, with
\begin{equation}
\label{def:mixture_model}
r_j \sim p^* \cdot \mathcal{U}\{u^*, \dots, m^*\} + (1-p^*)\cdot\mathcal{U}\{m^*+1, \dots, w^*\}, \quad1\le  u^* \le m^* < w^* \le n.
\end{equation}
All parameters---the boundary points $u^*$ and $w^*$, $p^* \in (0, 1)$, and $m^*$---are unknown and must be estimated from the data. In this framework, $m^*$ defines the location of the structural break in terms of rank-normalized intensity. The first component, $\mathcal{U}\{u^*, \dots, m^*\}$, represents the ``friendly'' interactions, while the second component, $\mathcal{U}\{m^*+1, \dots, w\}$, captures the background noise. The parameter $p^*$ reflects the mixture weights. Consequently, the probability of observing each a rank $r_j$ is 
\[
p(r_j) := 
\begin{cases} 
\frac{p^{*}}{m^* - u^* + 1}, & \text{if } r_j \le m^*, \\
\frac{1 - p^{*}}{w^* - m^*}, & \text{if } r_j > m^*.
\end{cases}
\]
Following this model, we define the set of ``friends'' as $F :=\, \{c_j \in C:\, r_j \le m^*\}$. The log-likelihood of the mixture model~\eqref{def:mixture_model}is
\[
L(p, m, u, w) := s\ln\left(\frac{p}{m-u+1}\right) + (k-s)\ln\left(\frac{1-p}{w - m}\right),
\]
where $s := \#\{j : r_j \le m\}$ denotes the number of ranks not exceeding $m$.

Let $r_{(1)} \le \dots \le r_{(k)}$ denote the ordered ranks. We estimate the boundaries as $\hat{u} := r_{(1)}$ and $\hat{w} := r_{(k)}$.
For fixed $(m,u,w)$, the maximization of $L$ with respect to $p$ yields
$\hat{p} = s/k$.
We then perform a discrete search over $m$ to obtain the maximizer $\hat{m}$.
The estimated set of ``friends'' is $\hat{F} = \{c_j \in C : r_j \le \hat{m}\}.$ In practice, the size of $\hat{F}$ may be sufficiently large (of order $k$). We discuss the interpretation of this case in Section~\ref{sec:interpr}.
Moreover, Section~\ref{sec:related_works} discusses alternative approaches to modeling and detecting structural breaks. 

\subsection{Detecting friendship}


To filter out the entities that do not exhibit ``friendship'', we check whether the ranks in the corresponding normalized interaction profile $\mathrm{row}_i(R)$ are distributed evenly across the observed range, i.e., for all $j$ 
\[
r_{j} \sim \mathcal{U}\{u^*, \dots, w^*\}, \quad 1\le u^* \le w^* \le n,
\]
where $u^*$ and $w^*$ are unknown parameters.  

To distinguish a true structural break from fluctuations, we propose two alternatives.

The first approach is the \textit{pre-fitting uniformity test}. That is, before estimating the mixture parameters, we assess whether the ranks in $\mathrm{row}_i(R)$ are uniformly distributed across the observed range $[\hat{u}, \hat{w}]$, where $\hat{u} = \min(r_j)$ and $\hat{w} = \max(r_j)$. While the use of empirical extrema for scaling introduces a conservative bias in the $p$-value estimation, this methodological aspect is justified since the priority is the identification of highly pronounced ``friendship'' patterns.

As an alternative to a uniformity test, we introduce an \textit{Information Criterion} that incorporates prior knowledge about the dataset. Suppose \textit{a priori} that the entity in hand $t_i$ has ``friends'' with probability $q \in (0, 1)$. We define two competing log-likelihoods:
\begin{equation}
L_1 := L(\hat{p}, \hat{m}, \hat{u}, \hat{w}) + \ln(q),
\quad
L_2 := k \ln\left(\frac{1}{\hat{w} - \hat{u} + 1}\right) + \ln(1 - q),
\end{equation}
where $L_1$ represents the log-likelihood under the structural break model (assuming $t_i$ has ``friends''), and $L_2$ corresponds to the model where $t_i$ has no ``friends''. The model with the higher value, $\max\{L_1, L_2\}$,  is selected.

\begin{algorithm}[ht!]
\SetAlgoLined
\DontPrintSemicolon
\KwIn{Interaction matrix $A \in \mathbb{R}^{n \times k}$, entity index $i$, prior probability $q \in (0,1)$, significance level $\alpha$, testing mode $M \in \{\text{Test, IC}\}$}
\KwOut{Estimated set of friends $\hat{F}$}

\BlankLine
\tcp{Step 1: Rank-based Representation}
For each column $\mathrm{col}_j(A)$, compute ranks $r_{ij}$ using randomized tie-breaking \;
Extract normalized profile $\mathrm{row}_i(R) = (r_{i1}, \dots, r_{ik})$ \;
Set $\hat{u} = \min(\mathrm{row}_i(R))$ and $\hat{w} = \max(\mathrm{row}_i(R))$ \;
Sort ranks such that $r_{(1)} \le r_{(2)} \le \dots \le r_{(k)}$ \;

\BlankLine
\tcp{Step 2: Pre-fitting Uniformity Check}
\If{$M = \text{Test}$}{
    $p_{val} \leftarrow \text{UniformityTest}(\mathrm{row}_i(R), [\hat{u}, \hat{w}])$ \;
    \If{$p_{val} > \alpha$}{
        \Return{$\emptyset$} \tcp*[r]{No structural break detected}
    }
}

\BlankLine
\tcp{Step 3: Maximum Likelihood Estimation}
$L_{max} \leftarrow -\infty$\;
\For{$m \in \{r_{(1)}, \dots, r_{(k-1)}\}$}{
    $s \leftarrow \sum_{j=1}^k \mathbb{I}(r_{ij} \le m)$ \;
    $\hat{p} \leftarrow s/k$ \;
    $L_{curr} \leftarrow s \ln\left(\frac{\hat{p}}{m-\hat{u}+1}\right) + (k-s) \ln\left(\frac{1-\hat{p}}{\hat{w}-m}\right)$ \;
    \If{$L_{curr} > L_{max}$}{
        $L_{max} \leftarrow L_{curr}$\;
        $\hat{m} \leftarrow m$\;
    }
}

\BlankLine
\tcp{Step 4: Model Selection via Information Criterion}
\If{$M = \text{IC}$}{
    $L_{null} \leftarrow k \ln\left(\frac{1}{\hat{w} - \hat{u} + 1}\right)$ \;
    \If{$L_{max} + \ln(q) \le L_{null} + \ln(1-q)$}{
        \Return{$\emptyset$} \;
    }
}

\Return{$\hat{F} = \{c_j \in C : r_{ij} \le \hat{m}\}$} 
\caption{The \textsf{friends.test} procedure}
\label{alg:friends}
\end{algorithm}

\section{Experimental results}
\label{sec:experiments}
We developed a novel R package \textsf{friends.test}  that implements the functionality described above and is available at 
\url{https://github.com/favorov/friends.test}. To validate our \textsf{friends.test} method on the real-world data, we applied it to the previously published transcriptomic dataset (GSE112026) \cite{ando2019chromatin, guo2016characterization}. That dataset contained 47 human papillomavirus-positive head and neck squamous cell carcinoma (HPV+ HNSCC) and 25 normal uvulopharyngoplasty (UPPP) surgical specimens. The method was applied to the RSEM-normalized gene expression matrix to identify the friends.  To post-process the data and to interpret the results, we look for group-specific markers among the genes with friends. We say that a gene is a group-specific marker if it has ``friends'' in at least 25\% of the target group samples and has zero ``friends'' in the opposing group. Based on this criterion, we identified cancer markers (associated exclusively with cancer samples) and normal tissue markers (associated exclusively with normal samples).

\begin{figure}[htbp]
     \centering
     \begin{subfigure}[b]{0.48\textwidth}
         \centering
         \includegraphics[width=\textwidth]{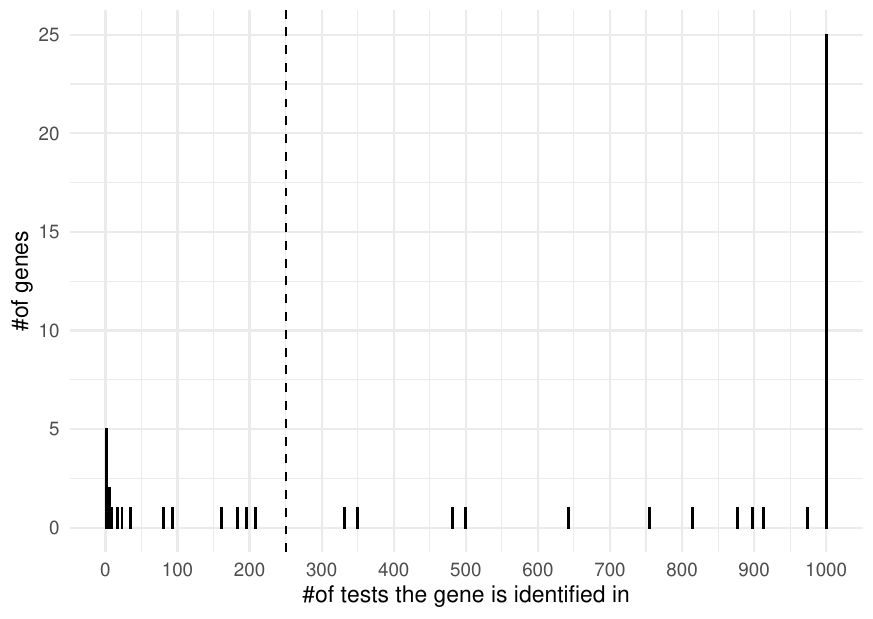}
         \caption{Permutation Test ($10^6$ iterations)}
         \label{fig:stability}
     \end{subfigure}
     \hfill
     \begin{subfigure}[b]{0.48\textwidth}
         \centering
         \includegraphics[width=\textwidth]{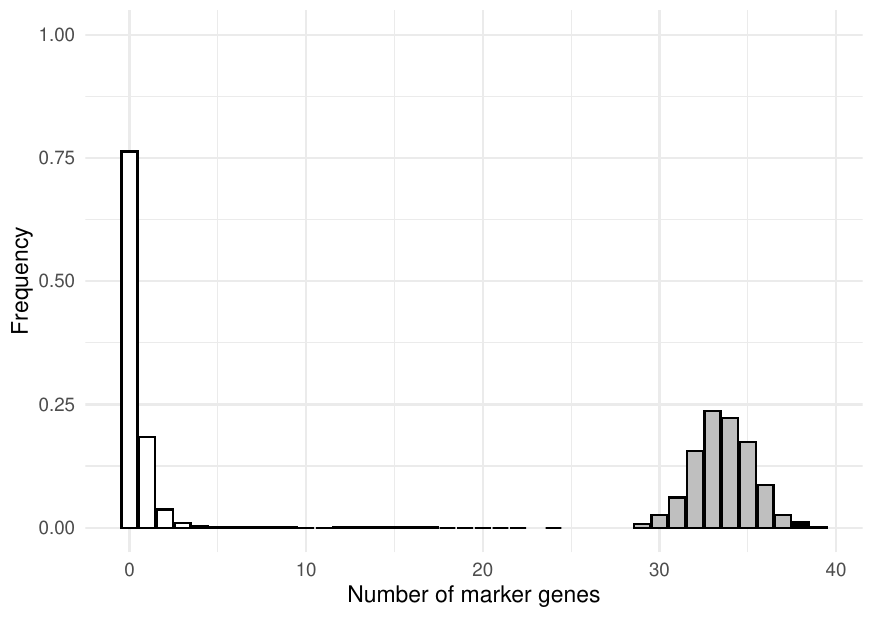}
         \caption{Stability Analysis ($10^3$ runs)}
         \label{fig:null_dist}
     \end{subfigure}

     \caption{Validation of the \textsf{friends.test} method. 
     (a) Frequency of gene identification across parallel runs, with the dashed line indicating the stability threshold (0.25). 
     (b) The empirical null distribution shows that the number of identified markers in the real data significantly exceeds random noise-based results. The white bars represent the $10^6$ permutations; the gray bars correspond to all the reliability test runs; the dark-gray bar consists the run that was user for permutations.}
     \label{fig:validation_plots}
\end{figure}

To ensure the reproducibility of the identified group-specific markers, we performed a stability analysis. The \textsf{friends.test} algorithm (with the consequent selection of genes and the group-specific markers) was executed $10^3$ times in parallel. The results of all iterations were aggregated to calculate the selection frequency for each gene. Genes identified in $>25\%$ of runs were retained as stable markers (Fig.~\ref{fig:stability}). This procedure yielded 37 markers: 35 cancer-specific and 2 normal-specific. Table~\ref{tab:stable_markers} presents the result. 

\begin{table}[h!]
\centering
\begin{tabular}{|l|p{12cm}|}
\hline
\textbf{Category} & \textbf{Stable Marker Genes} \\
\hline
Cancer Markers & ABCA13, AMDHD1, ATP13A5, C16orf73, C1orf110, CEL, COL11A1, COL22A1, COL7A1, COMP, CR2, CSAG2, CXorf22, CXorf59, CYP26A1, HOXD11, KRT17, LST-3TM12, MMP10, MMP13, MMP3, NKX2-4, NOS2, OCA2, PIWIL2, POSTN, PPP4R4, PRAME, PTH2R, SCUBE3, SLCO1B3, SOX14, SULT1E1, SYCP2, TG\\
\hline
Normal Tissue Markers & CLIC3, CR2 \\
\hline
\end{tabular}
\caption{List of stable markers identified for Cancer and Normal tissues. The annotation accords to GSE112026.}
\label{tab:stable_markers}
\end{table}

To assess the statistical significance of the identified group-specific markers, we performed a permutation test ($10^6$ iterations) by randomly shuffling group labels. The number of group-specific markers was compared with the number of markers in the permutation-based lists (Fig.~\ref{fig:null_dist}). All of the permutation-based lists were shorter than the group-specific marker lists. This result confirmed that the signal was stronger than random noise with $p$-value $< 10^{-6}$.

To study biological functions of 37 stable markers and the their relationship to cancer, we performed a literature analysis (see Section~\ref{sec:functions_of_genes}), and found that the 35 cancer markers collectively describes an invasive and remodeling tumor phenotype, including the invasion machinery (\textit{MMP3}, \textit{MMP10}, \textit{MMP13}) and extracellular matrix (ECM) (\textit{POSTN}, \textit{SCUBE3}, cancer-associated fibroblast (CAF) markers \textit{COL11A1}, and \textit{COL22A1}), as well as tumor-specific antigens (\textit{CSAG2, PRAME, KRT17, SYCP2}). The two normal markers included Chloride Intracellular Channel 3 (CLIC3)  and complement C3d receptor 2 (CR2). These results confirm that our \textsf{friends.test} method functions as a high-fidelity biological filter. Section~\ref{sec:functions_of_genes} provides a full list of genes' functions. Additionally, we have performed GSEA-MSigDB enrichment analysis for the  C2 gene set collection \cite{msigdb_genesets} of the cancer marker genes. This analysis showed that those genes were overrepresented mainly in extracellular matrix (ECM) remodeling pathways and in the HNSCC early markers set (see Supplementary File 1).

The experimental results confirm that the algorithm functions as a biological filter. By isolating these 37 genes, the method successfully recovered the core pathology of HNSCC: the loss of normal lymphoid structure (\textit{CR2}), the acquisition of invasive capability (\textit{MMPs}, \textit{POSTN}), and the restructuring of the tumor microenvironment (\textit{COL11A1}).
Similarity analysis of cancer marker genes, based on their sets of friends, revealed the underlying functional structure of the gene set.

\subsection{Friends' set similarity analysis}
\label{sec:discuss_res} 
\begin{figure}[ht!]
    \centering
    \includegraphics[width=0.5\textwidth]{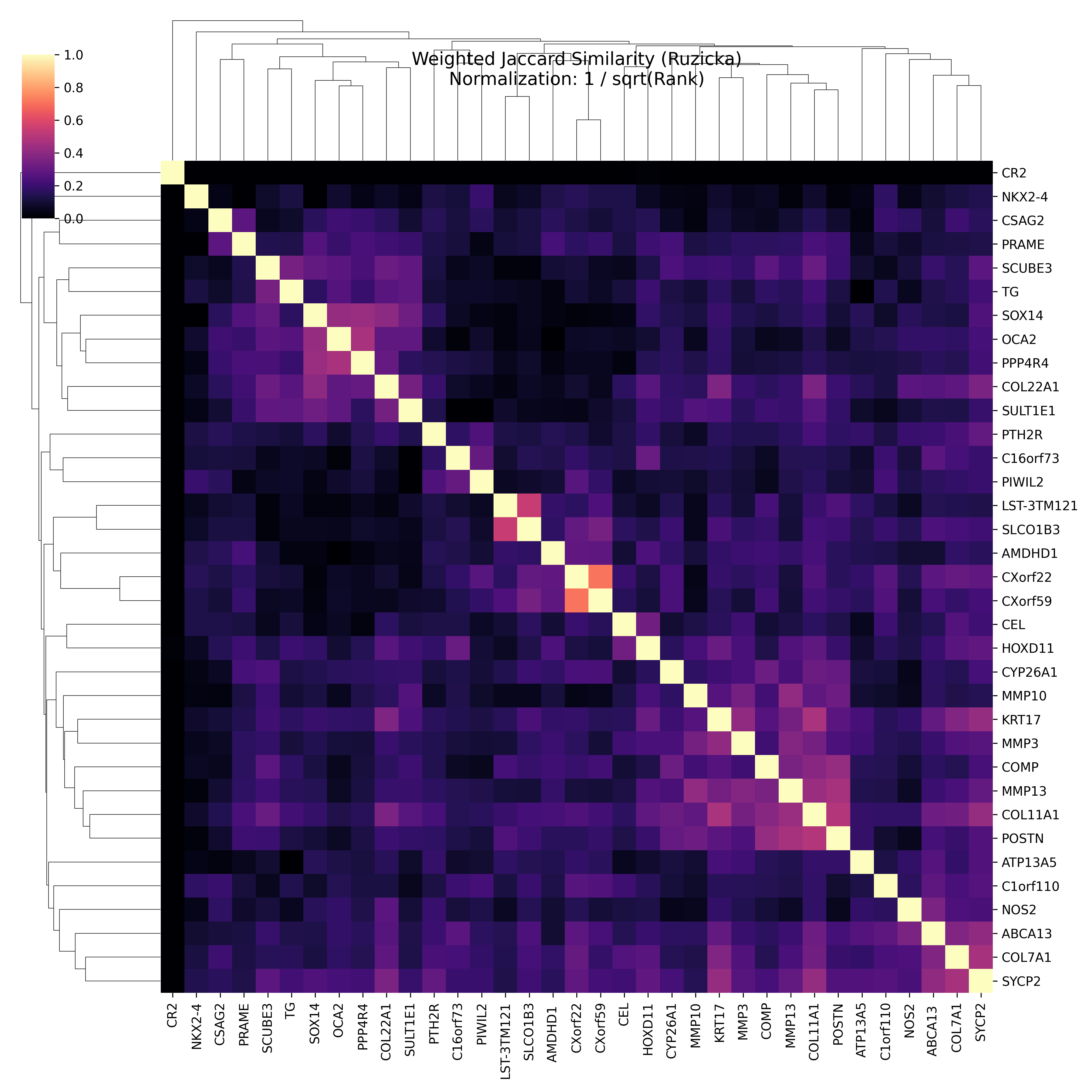} 
    \caption{Hierarchical tree based on Weighed Jaccard Similarity.}
    \label{fig:clustering}
\end{figure}

To illustrate the applicability of the method for assessing functional similarity between cancer gene markers, we utilized the Weighted Jaccard Similarity (also known as Ruzicka Similarity) \cite{cha2007comprehensive}. First, we constructed a global feature space defined by the union of all samples identified across all cancer marker genes. Each gene was then represented as a high-dimensional vector within this space. To characterize these marker genes, we utilized rank-based weighting. Specifically, for a given gene, each sample was assigned a weight based on its rank $r$ (defined as $r^{-1/2}$). Samples without an associated marker were assigned a weight of zero. Next, we quantified the pairwise functional overlap between genes using the Weighted Jaccard index. The resulting similarity matrix served as the input for hierarchical clustering (using the average linkage method \cite{virtanen2020scipy}), see Fig.~\ref{fig:clustering}. Section~\ref{sec:discuss_res} discusses the result.

The weighted Jaccard distance matrix (see Fig.~\ref{fig:clustering}) contains two pairs of very close genes. The first (and the most close) pair is \textit{CXorf59} and \textit{CXorf22}. Indeed, the up-to-date gene annotation has unified these identifiers under the gene symbol \genecard{CFAP47} (see \cite{ncbi_gene_170063}). The second pair is \textit{LST-3TM12} and \textit{SLCO1B3}. \textit{LST-3TM12} is a legacy identifier for a transcript now classified within the \textit{SLCO1B7} genomic region, involved in the same \textit{SLCO1B} gene family. The two genes are located adjacent on chromosome 12. Moreover, these two genes are sometimes transcribed in the same frame, forming a readthrough transcript protein  \cite{uniprot_F5H094}. Notably, the \textsf{friends.test} identifies specific interactions based solely on the internal structure of the input matrix, without relying on external biological databases or pre-existing gene annotations.

Based on the silhouette score, the hierarchical tree was pruned to yield nine clusters (Table~\ref{tab:gene_clusters} in Supplement). The functional identity of the clusters was subsequently verified through the same enrichment analysis procedure as for all 35 cancer marker genes. Only cluster 6 (see Table~\ref{tab:gene_clusters}) was overrepresented in any gene set of the C2 collection. Remarkably, now the HNSCC early markers set is the head of the list (see Supplementary File 2).

\section{Discussion}
\label{sec:discussion}

In this work, we introduced \textsf{friends.test}, an unsupervised approach designed to identify specific associations within bipartite interaction data. Our approach is motivated by the need to detect entities that exhibit high discriminative power---those that interact mostly with only a limited subset of counterparts, rather than exhibiting broad, non-specific connectivity across the entire dataset. Applying the method to head and neck squamous cell carcinoma transcriptomic data showed that the approach identified a small, stable set of differentially expressed genes.

\subsection{Related works}
\label{sec:related_works}
 Note that the interaction matrix $A$ inherently represents an adjacency matrix of a weighted bipartite graph (network), where two distinct sets of nodes are connected by edges representing their interaction strength. In network analysis, a rich family of methods focuses on identifying globally important nodes (hubs). These approaches prioritize broadly connected entities and capture global importance within a graph structure \cite{ma2017multi, tomasi2011functional}. As interaction datasets grow in complexity, graph anomaly detection has emerged as a critical field. It aims to identify unusual graph instances (nodes, edges, or subgraphs) that deviate significantly from the norm \cite{qiao2025deep, kim2022graph}. A significant area of research involves identifying dense sub-matrices or ``blocks'' within the interaction matrix, which allows for detecting structural changes at the matrix or submatrix level. Specifically, biclustering methods focus on discovering sub-matrices that satisfy specific homogeneity and statistical significance criteria \cite{castanho2024biclustering,jose2022biclustering, pontes2015biclustering}.

Our approach operates at the individual profile level (i.e., row-wise), rather than attempting to partition the entire interaction matrix at once. However, even at the row level, the choice of method remains critical. While clustering techniques like $k$-means are widely used for partitioning data, their application in our case presents some limitations. Specifically, $k$-means assumes relatively balanced cluster sizes and can be sensitive to outliers \cite{zhou2024adaptive , klochkov2021robust}, which may be the case when the number of ``friendly'' interactions and background interactions differ significantly. 

Similarly, change-point detection and segmentation methods, while designed to identify structural breaks, encounter difficulties when the selective signal involves only a few interactions. In such sparse scenarios, these methods may dismiss ``friendly'' interactions as outliers rather than indicate them as a meaningful structural shift.

Widely used specificity indices, such as the Tau index or the Gini coefficient, quantify how unevenly values are distributed within an interaction profile, thereby providing a single-number measure of overall specificity. However, these metrics do not identify which particular interactions constitute the specific signal \cite{kryuchkova2017benchmark}.

\subsection{Limitations of the approach}
\label{sec:limitations}
The proposed method is designed to detect specific interactions. In particular, we assume that specificity manifests as a separation between a relatively small set of ``friendly'' interactions and background ones. Profiles in which interaction strength changes gradually, or where the signal is distributed smoothly across many counterparts, may not exhibit a well-defined separation and can therefore be classified as non-specific. However, the experiment demonstrates that the method performs well on real data, where the model can be misspecified.

The use of empirical extrema for rank scaling in the uniformity test (Step 2 in Algorithm~\ref{alg:friends}) introduces a potential bias in $p$-value estimation toward conservatism. This approach may lead to the exclusion of genes with moderately expressed structural breaks. However, this conservative filtering is justified within the framework of identifying highly-specific interactions.

Moreover, an additional source of variability arises from the randomized tie-breaking procedure. While this approach prevents systematic bias, it can introduce fluctuations in the estimators, particularly when interaction profiles contain many ties or when the signal is weak. For this reason, we recommend running the procedure multiple times and assessing the stability of the identified ``friends'' across runs.

Finally, the proposed framework explicitly assumes an asymmetric interaction structure. When the interaction matrix $A$ is symmetric, this assumption breaks down, and the method is not guaranteed to perform well.

\subsection{Possible applications and interpretation of the results}
\label{sec:interpr}

\paragraph{Feature selection and graph sparsification.} Our approach serves as a tool for feature selection. In high-dimensional datasets, identifying a small subset of ``informative markers''---entities in $T$ that exhibit a distinct ``friendship'' pattern---allows for dimensionality reduction without losing the structural essence of the dataset. Furthermore, applying the friendship model to complex interaction networks enables principled sparsification of bipartite graphs. Instead of working with a dense, noisy adjacency matrix, we retain only edges corresponding to the ``friendly'' interactions.

\paragraph{Functional similarity and clustering.} As we have demonstrated, the concept of ``friendship'' provides a basis for guilt-by-association paradigms: two entities might be considered as functionally similar if they share significantly overlapping sets of ``friends''. Moreover, the introduction of functional similarity leads to an interpretable clustering. 

\paragraph{``Anti-friends'' and negative selection.} The current model identifies strong positive interactions. However, if the estimated set of ``friends'' $\hat{F}$ is large, while its complement is small, one may suppose a significant absence of interaction (the presence of ''anti-friends''). We do not explicitly model or validate such effects in the current study; however, these cases may motivate future extensions of the framework aimed at capturing inhibitory or mutually exclusive relationships.

\paragraph{Alternative functional shapes.} While the current implementation utilizes a step function to identify ``friends'', the underlying likelihood framework remains inherently flexible. However, future research could explore alternative shapes to better capture more complex data behaviors. For instance, bump functions could identify interactions that occur within a specific range of latent intensity.

\section{Acknowledgements}
The authors are grateful to Dr. Vasily Ramensky for fruitful discussions that significantly shaped the project along its history and to Dr. Anatoliy Rubinov for very constructive critique.
The authors acknowledge support by Break Through Cancer to DL, EJF and AF.

\bibliographystyle{unsrt}  

\bibliography{references}




\newpage
\appendix

\section{Supplementary material}
\label{sec:functions_of_genes}
\begin{small}
\begin{tabularx}{\textwidth}{|l|l|X|X|}
\hline
\textbf{\#} & \textbf{Gene} & \textbf{Function} & \textbf{Role in Cancer} \\ \hline
\endfirsthead
\hline
\textbf{\#} & \textbf{Gene} & \textbf{Function} & \textbf{Role in Cancer} \\ \hline
\endhead

1 & \genecard{ABCA13} & ATP-binding cassette transporter; lipid transport & Overexpressed in some tumors; linked to adhesion and angiogenesis regulation \\ \hline
2 & \genecard{AMDHD1} & Amidohydrolase in histidine catabolism & Tumor suppressor in cholangiocarcinoma; inhibits metastasis via TGF-b/SMAD pathway \\ \hline
3 & \genecard{ATP13A5} & P5-type ATPase; cation transport & No direct cancer role documented; expressed in some tumors \\ \hline
4 & \genecard{C16orf73} & Single-stranded DNA-binding; meiosis & No established cancer role \\ \hline
5 & \genecard{C1orf110} & Coiled-coil protein; cell cycle/DNA repair & No clear cancer role \\ \hline
6 & \genecard{CEL} & Pancreatic lipase; lipid metabolism & Altered in pancreatic cancer; may affect tumor metabolism \\ \hline
7 & \genecard{COL11A1} & ECM collagen; CAF marker & Invasion-supporting machinery \\ \hline
8 & \genecard{COL22A1} & ECM collagen; CAF marker & Invasion-supporting machinery \\ \hline
9 & \genecard{COL7A1} & Basement membrane collagen & Basement membrane remodeling; linked to invasion \\ \hline
10 & \genecard{COMP} & ECM glycoprotein & Promotes metastasis and poor prognosis in breast and prostate cancer \\ \hline
11 & \genecard{CSAG2} & Cancer/testis antigen & Immune evasion; biomarker in melanoma and sarcoma \\ \hline
12 & \genecard{CXorf22} & PRAME family & Immune escape; poor prognosis marker \\ \hline
13 & \genecard{CXorf59} & PRAME family & Immune escape; poor prognosis marker \\ \hline
14 & \genecard{CYP26A1} & Retinoic acid hydroxylase & Promotes proliferation, invasion, EMT; poor prognosis in multiple cancers \\ \hline
15 & \genecard{HOXD11} & Homeobox transcription factor & Promotes invasion and metastasis; poor prognosis in SCC \\ \hline
16 & \genecard{KRT17} & Type I keratin; cytoskeleton & Squamous cell identity; promotes growth and migration; poor prognosis \\ \hline
17 & \genecard{SLCO1B7} & Organic anion transporter variant & No direct cancer role documented \\ \hline
18 & \genecard{MMP10} & Matrix metalloproteinase; ECM degradation & Supports invasion, metastasis; poor survival in SCC \\ \hline
19 & \genecard{MMP13} & Collagenase; ECM remodeling & Promotes invasion and metastasis; poor prognosis \\ \hline
20 & \genecard{MMP3} & Stromelysin; ECM degradation & Facilitates invasion and angiogenesis; aggressive phenotype \\ \hline
21 & \genecard{NKX2-4} & Homeobox transcription factor & Reported in EMT and stemness; limited data \\ \hline
22 & \genecard{NOS2} & Nitric oxide synthase & Promotes angiogenesis and tumor progression; context-dependent \\ \hline
23 & \genecard{OCA2} & Melanosome pH regulator & No strong cancer link; pigmentation biology \\ \hline
24 & \genecard{PIWIL2} & piRNA pathway protein & Oncogenic; promotes stemness and resistance to apoptosis \\ \hline
25 & \genecard{POSTN} & ECM protein; cell adhesion & Promotes invasion, metastasis, angiogenesis; poor prognosis \\ \hline
26 & \genecard{PPP4R4} & PP4 regulatory subunit & DNA repair and cell cycle; limited cancer data \\ \hline
27 & \genecard{PRAME} & Cancer-testis antigen & Immune evasion; poor prognosis marker \\ \hline
28 & \genecard{PTH2R} & GPCR for parathyroid hormone & Minimal cancer data; possible microenvironment signaling \\ \hline
29 & \genecard{SCUBE3} & Secreted glycoprotein & Promotes proliferation, EMT, metastasis; poor prognosis \\ \hline
30 & \genecard{SLCO1B1} & Organic anion transporter & Pharmacogenomic relevance; no strong cancer role \\ \hline
31 & \genecard{SLCO1B3} & Organic anion transporter & Overexpressed in some cancers; drug resistance link \\ \hline
32 & \genecard{SOX14} & Transcription factor & May promote EMT and stemness; limited data \\ \hline
33 & \genecard{SULT1E1} & Estrogen sulfotransferase & Alters estrogen signaling; implicated in hormone-dependent cancers \\ \hline
34 & \genecard{SYCP2} & Synaptonemal complex protein & Cancer-testis antigen \\ \hline
35 & \genecard{TG} & Thyroglobulin precursor & Marker for thyroid cancer; used clinically for monitoring \\ \hline
36 & \genecard{CLIC3} & Growth regulator & Down-regulated in HNSCC \\ \hline
37 & \genecard{CR2} & Interface between innate and adaptive immune systems & \\ \hline
\end{tabularx}
\end{small}

\begin{table}[ht!]
\centering
\caption{Hierarchical clustering split into $k=9$ clusters}
\label{tab:gene_clusters}
\begin{tabularx}{\textwidth}{l c X}
\toprule
\textbf{Cluster} & \textbf{Number of marker genes} & \textbf{List of marker genes} \\ 
\midrule
Cluster 1 & 2 & CSAG2, PRAME \\ \addlinespace
Cluster 2 & 7 & COL22A1, OCA2, PPP4R4, SCUBE3, SOX14, SULT1E1, TG \\ \addlinespace
Cluster 3 & 3 & C16orf73, PIWIL2, PTH2R \\ \addlinespace
Cluster 4 & 5 & AMDHD1, CXorf22, CXorf59, LST-3TM121, SLCO1B3 \\ \addlinespace
Cluster 5 & 2 & CEL, HOXD11 \\ \addlinespace
Cluster 6 & 8 & COL11A1, COMP, CYP26A1, KRT17, MMP10, MMP13, MMP3, POSTN \\ \addlinespace
Cluster 7 & 6 & ABCA13, ATP13A5, C1orf110, COL7A1, NOS2, SYCP2 \\ \addlinespace
Cluster 8 & 1 & NKX2-4 \\ \addlinespace
Cluster 9 & 1 & CR2 \\ 
\bottomrule
\end{tabularx}
\end{table}

Further, we summarize the information about the identified cancer marker genes,
\begin{itemize}
    \item \textit{COL11A1} is widely validated as a specific marker for CAFs in the head and neck tumor microenvironment \cite{nallanthighal2021collagen, garcia2013overexpression}.
    \item \textit{MMP3}, \textit{MMP10}, and \textit{MMP13} are critical for degrading the basement membrane, facilitating tumor invasion. Specifically, \textit{MMP10} and \textit{MMP13} are known to correlate with metastasis and poor survival in HNSCC \cite{deraz2011mmp, luukkaa2006association, vincent2014overexpression, iizuka2014matrix}.
    \item  \cite{wang2023machine} identified COL7A1 as a top-ranking diagnostic predictor specifically for squamous cell carcinomas, including Head and Neck Squamous Cell Carcinoma (HNSC) and Lung Squamous Cell Carcinoma (LUSC).
    \item \textit{POSTN} (Periostin) 
    functions as a hub gene for cell adhesion and migration, bridging cancer cells with the structural matrix \cite{teng2025periostin, zhao2022hub, xian2025periostin}.
    \item Secretory SCUBE3 supports oncogenic activity through interactions with key oncogenic cell surface receptor proteins \cite{singh2025antibody}.
    \item \textit{PRAME} is highly specific to HNSCC and melanoma and is associated \cite{szczepanski2013prame} with retinoid resistance \cite{ramchatesingh2025targeting}
    \item \textit{KRT17} is identified as a critical mediator of drug resistance and immune evasion in HNSCC \cite{hou2025krt17}.
    \item  \textit{SYCP2} is expressed in HPV associated Cancers \cite{almubarak2021investigating}.
\end{itemize}

\vspace{1in}
Supplementary file 1: The GSEA MSigDB result for the cancer marker genes. \\ \href{https://raw.githubusercontent.com/favorov/friends-test-manuscript-code/refs/tags/arxiv_1.0/RNAseq_GSE112026/results/GSE112026/GSEA_MSigDB_cancer_genes.tsv}{GSEA\_MSigDB\_cancer\_genes.tsv}

Supplementary file 2: The GSEA MSigDB result for the cluster 8 cancer marker genes. \\ \href{https://raw.githubusercontent.com/favorov/friends-test-manuscript-code/refs/tags/arxiv_1.0/RNAseq_GSE112026/results/GSE112026/GSEA_MSigDB_cluster_6_8_cancer_genes.tsv}{GSEA\_MSigDB\_cluster\_6\_8\_cancer\_genes.tsv}



\end{document}